\documentstyle[12pt]{article}
\begin{document}


\vskip 5mm

\begin{center}
{\large \bf
A Relativistic   Approach  to \\
  Deep Inelastic Scattering on the Deuteron.
}\\[4mm]
\noindent
  A.Yu. Umnikov,
L.P. Kaptari\footnote{On leave from \em
 Bogoliubov's Laboratory of Physics, JINR,
Dubna, 141980 Russia},
K.Yu. Kazakov\footnote{On leave from \em
 Far Eastern State University, Vladivostok, 690000 Russia}
 and F.C. Khanna\\[4mm]
{\em
 Theoretical Physics Institute, Physics Department,
 University of
 Alberta, \\Edmonton,
  Alberta T6G 2J1,  and \\
 TRIUMF, 4004 Wesbrook Mall, Vancouver, B.C. V6T 2A3, CANADA}
 \end{center}

\begin{abstract}
A  covariant field theoretical approach to
deep inelastic scattering on the deuteron is presented.
The  deuteron structure function
is calculated in terms of the Bethe-Salpeter amplitude.
Numerical
calculations
for the
nucleon contribution
 are made with  a realistic model of the $NN$-interaction, including
$\pi$-, $\rho$-, $\omega$-,
 $\eta$-, $\delta$- and $\sigma$-mesons, and results
 are compared with previous
  non-relativistic calculations.
\end{abstract}


\newpage

 {\bf 1. Introduction}

 The relativistic bound state problem appears as one of the most
 important and
 interesting in the modern theory of strong interactions.
  Since the bound state problem in QCD is still unresolved,
 it is important to study the capability of
 the effective theories both in the quark and hadron
 sectors.
 The most direct way to describe
 the bound states in a field theory
 is to consider the
 Bethe-Salpeter (BS) equation~\cite{reviewb}. Approaches
 based on the
 BS equation,  or its approximations,
 within models using effective potentials,
 allow for a successful description of the mass spectrum and the
 decay widths of
 mesons as bound systems of two quarks
(for recent development see e.g. refs.~\cite{qqnew}).
 The BS equation has also been applied
 to describe properties of the deuteron~\cite{tjond}, and
 effective NN-forces~\cite{gourdin,gross}.

 This letter presents a
 relativistic approach to deep
 inelastic lepton scattering on the deuteron based on the BS
 equation. This process was previously studied on the
 basis of various relativistic equations~\cite{fs}-\cite{meln},
 approximations to the BS
 equation, and the estimates
 of the nuclear effects in the deuteron structure function (SF)
$F_2^D$
 display quite strong dependence on the approximation are made.
 We present a
 consistent covariant description of the
 nuclear effects in SF of
 the deuteron based on the BS equation
 with a realistic $NN$-interaction, including
 $\pi$-, $\rho$-, $\omega$-,
 $\eta$-, $\delta$- and $\sigma$-mesons. We calculate $F_2^D$
 in terms of the BS amplitude, using the operator
 product expansion method
 within the effective meson-nucleon theory.
 More details about formalism may be found in ref.~\cite{umkh}.

 Our investigation is  motivated partly
 by a number of  existing and forthcoming
 experiments on deep inelastic scattering
 of leptons by deuterons (SLAC,  CERN, DESY, CEBAF).
 The points of interest are (i)  to
 extract the neutron SF from the combined proton-deuteron data
 and (ii) to estimate the ``EMC-effect" on the deuteron, combining
 the electroproduction and neutrino data~\cite{bodsmi}
(c.f.~\cite{fs,gom}).
 A relativistic theory of this process will
 be beneficial  in the
 analysis of the experimental data.

{\bf 2. Deep inelastic scattering on
the deuteron in the Bethe-Salpeter formalism}

 The
 hadronic tensor, $W_{\mu \nu}$,  is proportional to
 the imaginary part of the
 virtual photon-hadron elastic scattering amplitude:
   \begin{eqnarray}
  T_{\mu \nu}(P, q)
  = {\it i}\int d^4x \, e^{{\it i} qx}
  \langle  P \mid  T(J_{\mu}(x)J_{\nu}(0))
  \mid P \rangle,
  \label{f03}
  \end {eqnarray}
where $P$ is
the
momentum of the target and $q$ is the virtual photon momentum
($Q^2 \equiv - q^2$).
 Thus the calculation of $W_{\mu \nu}$ is  split into two
relatively independent parts: a description of the bound state,
and of the operator to be sandwiched between this state.

An accurate description
of both the $NN$-interaction at
 energies up to $\sim$1 GeV, and the basic
 properties of the deuteron, can be provided within the
 meson-nucleon theory~\cite{tjond,gross}.
 The covariant description is
 based on the BS equation or its various
 approximations.  We use
 the ladder approximation for the kernel of the
  BS equation~\cite{tjond}:
  {\small
\begin{eqnarray}
\Phi(p, P_D) =
i \hat S(p_1)\cdot\hat S(p_2) \cdot
\sum_{B} \int \frac{d^4p'}{(2\pi)^4}
\cdot
\frac{g^2_B \Gamma_{B}^{(1)}
\otimes
\Gamma_{B}^{(2)}}{(p-p')^2-\mu_B^2}
\cdot \Phi(p', P_D),
\label{sseq}
\end{eqnarray}
}
where  $\mu_B$ is the mass
 of meson $B$;
$\Gamma_{B}$ is the meson-nucleon vertex,
corresponding to the meson $B$, $\hat S(p) = (\hat p - m)^{-1}$,
$m$ is the nucleon mass.
The amplitude  $\Phi$ appears as a
$4\times 4$-matrix in the spinor space,
so in the most general case the two-spinor BS amplitude consists of
 16 independent components.
However, due to the parity invariance of the strong interactions,
these amplitudes are split into two independent sets with different
parities. Only eight components, with a positive parity
are relevant to describe  the deuteron.
Decomposing  $\Phi$ in  terms of the
complete set of the
Dirac matrixes, their bilinear combinations
 and  the $4 \times 4$-identity matrix, $\hat{\sf 1}$,
and performing the partial wave analysis we obtain
for the deuteron a
 system of eight coupled two-dimensional
integral equations with singular kernels.
 To remove the singularities from the kernels
we use the well-known Wick rotation (see, e.g.~\cite{reviewb}
and references therein). More technical
details may be found in~\cite{umkh}.

The meson parameters, such as
masses, coupling constants, cut-off parameters
are taken similar to those  in ref.~\cite{tjond},
with a minor adjustment
of the coupling constant of the scalar $\sigma$-meson
 so as to provide a numerical solution of the BS
equation.
 All parameters are presented
in Table 1, where coupling constants are shown in accordance with
our definition of the meson-nucleon form-factors, $F_B(k) =
(\Lambda^2 -
\mu_B^2)/(\Lambda^2 - k^2)$.

 As to the second part of the problem, to describe
 the interaction operator in (\ref{f03})
 we utilize the Wilson Operator Product Expansion (OPE) method
 within the same
 meson-nucleon theory~\cite{levin,physlet,umkh}.
  The OPE of the spin-averaged Compton
   amplitude (\ref{f03}) has the form~\cite{r1}:
 {\small
  \begin{eqnarray}
T_{\mu \nu}^D(P_D, q)=
\! \sum^{\infty}_{a; n=2,4, \ldots }\! \! C^{(1)}_{a, n}
\left(-g_{\mu\nu}
 + \frac{q_{\mu}q_{\nu} }{q^2}
 \right)
 \frac{ 2^nq_{\mu_1} \ldots q_{\mu_{n}} } { {(-q^2)}^n }
 \langle  P_D\mid{\sl O}^{\mu_1 \ldots \mu_n}_{a}(0)\mid P_D \rangle
 \label{f04}
 \end {eqnarray}
	\vspace*{-.6cm}
 \begin{eqnarray}
+\! \sum^{\infty}_{a; n=2,4, \ldots }\! \! C^{(2)}_{a, n} \
  \left(g_{\mu \mu_1} - \frac{q_{\mu}q_{\mu_1} }{q^2} \right)
  \left(g_{\nu \mu_2} - \frac{q_{\nu}q_{\mu_2} }{q^2} \right)
 \frac{2^nq_{\mu_3} \ldots q_{\mu_n} } { {(-q^2)}^{n-1}}
 \langle  P_D\mid{\sl O}^{\mu_1 \ldots \mu_n}_{a}(0)\mid P_D\rangle,
  \nonumber
  \end {eqnarray}
  } 
 where ${\sl O}^{\mu_1 \ldots \mu_n}_{a}(0)$ is a set of
 local operators providing the basis for the OPE;
 $a$ enumerates operators of different fields of the theory;
 $C^{(1,2)}_{a, n}$ are the Wilson's coefficient functions.
The structure functions, e.g. $F_2$, are obtained by
 using
 the dispersion technique~\cite{r1} and is presented in the form:
\begin{equation}
{\sf M}_{n-1}(F_{2}^D) = \sum_{a} C^{(2)}_{a,n} \cdot
 \mu^{a/D}_{n},
  \quad
   {\rm with }
     \quad
       {\sf M}_n(F) = \int
         \limits_0^1 F(x) x^{n-1} dx,
          \label{f307}
           \end{equation}
where $x$ is the Bjorken scaling variable, $0 \leq x \leq 1$ and
$\mu^{a/D}_{n}$ is defined by:
   \begin{equation}
     \langle P_D \mid {\sl O}_a^{\mu_1 \ldots \mu_n} \mid P_D \rangle
\equiv
      \mu^{a/D}_{n} \cdot P^{\mu_1}_D \ldots P^{\mu_n}_D ,
         \quad (a=N,B ).
         \label{f066}
          \end{equation}

To obtain the moments of the deuteron
SF (\ref{f307}) in the meson-nucleon theory we, first,
define the set ${\sl O}^{\mu_1 \ldots \mu_n}_{a}(0)$.
For deep inelastic scattering the leading operators,
twist two,  for nucleon fields have
 the form (symmetrisation and subtraction of traces
are assumed):
\begin{eqnarray}
{\sl O}_{\psi}^{\mu_1 \ldots \mu_n} = \frac{1}{2}\left( \frac{ {\it
i} }{2}
\right)^{n-1}
 \psi (0)\gamma^{\mu_1}
\!\!\stackrel{\leftrightarrow}{\partial}
\! \! \! \phantom{!}^{\mu_2}\!\ldots
\stackrel{\leftrightarrow}{\partial}
\! \! \! \phantom{!}^{\mu_n} \psi (0).
\label{211}
\end{eqnarray}
The operators for meson fields are similar to (\ref{211})
(c.f.~\cite{levin,physlet,umkh}).
Secondly, the coefficients,  $C^{(1,2)}_{a, n}$,
are target-independent and
by consideration of the scattering
on the free nucleons and mesons,
we find that in the ladder approximation
 these coefficients are identical to
  moments of the SF $F_2^a$ of nucleons $(a=N)$
   or mesons $(a=B)$:
$ C^{(2)}_{N, n}  = {\sf M}_n(F_2^N) ,
\quad
 C^{(2)}_{B, n}= {\sf M}_n(F_2^{B})$.

{}From eqs. (\ref{f066})-(\ref{211})  we  get
the following form for moments of the deuteron
SF:
\begin{eqnarray}
 {\sf M}_n(F_2^D)=
{\sf M}_n(F_2^N) \cdot \mu^{N/D}_{n+1} + \sum_{B}{\sf M}_n(F_2^{B})
\cdot \mu^{B / D}_{n+1},
\label{md}
\end{eqnarray}
where $\mu^{N/D}_{n}$ and $\mu^{B/D}_{n}$  are interpreted as
  moments of the effective distribution functions
of the nucleons and mesons in the deuteron, respectively.
In the framework of
our approach the meson and nucleon
SF are considered as given
and moments
of effective distributions
are the matrix elements of the
twist-2 operators on the deuteron state, computed by
the Mandelstam method~\cite{mandelstam}. For the nucleon contribution
to
the moments of the deuteron structure function we get:
{\small
\begin{eqnarray}
\mu^{N/D}_{n} = \frac{1}{M_D^{n} }
 \int \frac{d^4p}{(2\pi)^4}
 {\bar \Phi_D(p, P_D)}\left \{
 \hat S^{-1}(p_2)
 \left ( \gamma_0^{(1)} + \gamma_3^{(1)} \right ) \cdot
\left ( p_{10}+p_{13} \right ) ^{n-1}\right \}\Phi_D(p, P_D)
\label{mun} \end{eqnarray}
 }
where  the kinematical variables in the
rest frame are defined by
 \begin{eqnarray}
p = (p_0,{\bf p}),
\quad p' = (p'_0,{\bf p'}),
\quad
P_D=(M_D, {\bf 0}), \quad
p_1 = \frac{P_D}{2} + p, \quad p_2 = \frac{P_D}{2} - p,
\label{kin}
\end{eqnarray}
where $M_D$ is the deuteron mass,
$p_{1k}$ is $k^{\rm th}$ component of
four-vector $p_1$ and in the deep inelastic kinematics
$pq \approx q_0 (p_0+p_3)$.

Using the explicit form of the moments
(\ref{md})-(\ref{mun})
and the  inverse Mellin transform,
we obtain the deuteron SF in the
convolution form:
{\small
\begin{eqnarray}
&& F_2^D(x) = \int \limits_{0}^{1} f^{N/D}(\xi) F_2^N (x/\xi)d\xi
\quad+ \quad
{ meson \quad exchange\quad terms},
\label{conv}\\
f^{N/D} (\xi) & = &\frac{1}{M_D }
 \int \frac{d^4p}{(2\pi)^4}
 {\bar \Phi_D(p, P_D)}
  \hat S^{-1}(p_2)
 \left ( \gamma_0^{(1)} + \gamma_3^{(1)} \right )
 \Phi_D(p, P_D) \label{fn} \\
 && \phantom{2222}\left \{ \theta (p_{1 0}+p_{1 3}) \, \delta \!\!
 \left ( \xi - \frac {p_{10}+p_{13} }{M_D}
 \right ) + \theta (-p_{10}-p_{1 3}) \,\delta\!\!
 \left ( \xi + \frac {p_{1 0}+p_{1 3} }{M_D}
 \right )
 \right \}.
\nonumber
\end{eqnarray}}
This formula has
clear and obvious physical interpretation. The first term
on the r.h.s. represents the contribution of the nucleons
to the full deuteron SF, $F_2^D$,
{\em the relativistic impulse approximation}.
The second term is the contribution of {\em the meson
exchange current}, which can be also presented in
convolution form~\cite{umkh}.

{\bf 3. Numerical results and discussion}

Solving exactly the  dynamical
problem for the deuteron, described by eq. (\ref{sseq}),
 and computing the deuteron structure
function, eqs.~(\ref{conv})-(\ref{fn}),
  we obtain  {\em realistic} estimates for the
nucleon contribution
 to the deuteron SF, $F_2^D$, including
off-mass-shell effects (also referred to as ``binding
effects") and effects of Fermi motion
in the deuteron. In  fig.~1 we present   our  results
in form of the ratio, $F_2^D(x)/F_2^N(x)$. The solid line
is the result of calculation with the BS approach.
We compare
our calculations with non-relativistic results
and calculation in the
light-cone kinematics.
The nonrelativistic
calculations are presented within two approaches
(c.f.~\cite{physlet}),
the $x-$rescaling model,
referred to in fig.~1 as ``nonrelativistic-I" and OPE-OBE model,
``nonrelativistic-II". Both these approaches describe fairly well
 deep inelastic scattering off heavy nuclei. It is seen that in
the relativistic model the effects of nuclear structure of the
deuteron is
larger in comparison with
that in the non-relativistic calculations. Apparently,
this difference is
due to a more consistent consideration of the off-mass-shell
behavior of bound  nucleons within the BS formalism. The nuclear
structure
effects in the deuteron are taken into account in the BS amplitude
through
the binding energy $\varepsilon_D$ and the dependence of the
amplitude on
the relative four-momentum of nucleons.
 As an illustration of the influence of  binding effects in the
deuteron
we present in  fig.~1 the calculation in the light-cone
kinematics (dotted line),
disregarding the binding effects in the deuteron.
In this case the ratio, $F_2^D(x)/F_2^N(x)$,
is consistent with 1 at $x<0.5$ and, due to Fermi motion, becomes
greater
 than 1 at $x>0.5$.
 A comparison of the our results with this approach
shows that  the relativistic off-mass-shell
 effects
lead to an ``EMC-effect" on the deuteron, similar to that
experimentally
observed for heavy nuclei and theoretically obtained
in non-relativistic approaches for the deuteron.
All approaches may be compared also by computing
$\mu_1^{N/D}$, which is interpreted as
the fraction of the total momentum of the deuteron carried by
nucleons~\cite{fs,levin,physlet,umkh}.
Defining $\delta_N$ as the deviation of the $\mu_1^{N/D}$ from unity,
we obtain
$\delta_N \approx 1\cdot 10^{-2}$ which is twice
the corresponding nonrelativistic result,
$\delta_N \approx 5\cdot 10^{-3}$. At the same time for the
light-cone
calculation we have $\delta_N = 0$.

All curves in fig.~1 are calculated with a rather
simple parametrization of the nucleon SF, $F_2^N$, used in previous
non-relativistic calculations~\cite{physlet}. To emphasize the
dependence
of the ratio, $F_2^D(x)/F_2^N(x)$,
 on the parametrization for $F_2^N$ we show in fig.~2
calculations with a realistic parametrization
of the nucleon structure function $F_2^N(x)$~\cite{amb}, which allows
a consideration of the $Q^2$-dependence of the ratio as well. Results
for
$Q^2=10~GeV^2$ and $90~GeV^2$ are presented by solid lines, 1 and 2
respectively.
The dashed line is the same as the solid line in fig.~1,
where the parametrization of $F_2^N(x)$  corresponds to
low $Q^2$ and   its behavior differs from the
previous parametrization as $x\to 1$ .
Figure~2 demonstrates that
the ratio $F_2^D(x)/F_2^N(x)$ is almost $Q^2$-independent and its
shape is sensitive to
 the parametrization of the nucleon structure function
as $x\to 1$.

The absolute value of the deuteron structure
function, $F_2^D$, as $x \to 1$ is shown in fig.~3 (solid line).
 The relativistic
structure function displays
a harder behavior of the ``tail" in the vicinity of $x\sim 1$
compared to the non-relativistic $x$-rescaling calculations (dashed
line).

A more detailed analysis of the relativistic effects
in the deuteron SF, including the spin-dependent SF,
will be presented elsewhere~\cite{my}.

{\bf 4.  Conclusions}

  We have presented a relativistic approach to the deep inelastic
 lepton-deuteron scattering based on the Bethe-Salpeter formalism
 within an effective meson-nucleon theory. In particular,
 \begin{itemize}
 \vspace*{-3mm}
\item
 The spinor-spinor Bethe-Salpeter equation for the deuteron
 is solved in the ladder approximation
 for a realistic meson exchange potential, including
 $\pi$-, $\rho$-, $\omega$-, $\eta$-, $\delta$- and $\sigma$-mesons.
 \vspace*{-3mm}
 \item
  The structure function $F_2^D$ of the deuteron is  calculated
 in terms of the Bethe-Salpeter amplitudes
 by the Operator Product
 Expansion method within the meson-nucleon theory.
 \vspace*{-3mm}
  \item
  Our numerical calculations of the structure function
  $F_2^D$ emphasize a qualitative agreement with previous
non-relativistic
   results,  the magnitude of the effects of nuclear structure
   have been found  to be larger in the relativistic approach.
 \end{itemize}
 \vspace*{-3mm}


{\bf 5. Acknowledgments}

 The research is supported in part
 by the NSERC (Canada).
 One of the author (L.P.K.) wishes to thank
NSERC for an international
scientific exchange award that made it possible for him
 to visit
the University of Alberta.


  \newpage

\begin{center}
 Table 1. The parameters of the model.

\begin{tabular}{|c|c|c|c|c|}
\hline
 meson & coupling constants       & mass     & cut-off & isospin   \\
 B   & $g^2_B/(4\pi); [g_t/g_v]$ &$\mu_B$, GeV  &$\Lambda$,GeV &
\\
 \hline
\hline
 $\sigma$ & 12.2 & 0.571 & 1.29 & 0  \\
\hline
 $\delta$ & 1.6 & 0.961 & 1.29  & 1 \\
\hline
 $\pi$ & 14.5 & 0.139 & 1.29   & 1\\
\hline
 $\eta$ & 4.5 & 0.549 & 1.29  & 0 \\
\hline
 $\omega$ & 27.0; [0] & 0.783 & 1.29  & 0 \\
\hline
 $\rho$ &  1.0; [6] & 0.764 & 1.29 & 1  \\
\hline
\hline
 \multicolumn{5}{|c|}{$m = 0.939$ GeV, $\epsilon_D = -2.225$ MeV}\\
\hline
\end{tabular}
\end{center}

\newpage

\vskip 2cm

\begin{center}
{\bf \LARGE Figure captions}
\end{center}

\vskip 2cm

%
%

Figure 1. {\em The ratio of the deuteron and nucleon structure
functions
$F_2^D(x)/F_2^N(x)$ calculated in different theoretical approaches.
Curves:
Solid  is the present result of  the BS formalism ;
Dashed and dot-dashed are previous non-relativistic estimates (see
text);
Dotted is calculation in the light-cone kinematics.
The nucleon SF, $F_2^N$ is taken from
ref.~{\rm \cite{physlet}}.}\\[1cm]

%
%
%

Figure 2. {\em The ratio
$F_2^D(x)/F_2^N(x)$ calculated
with different parametrization for $F_2^N$.
Solid  lines present calculations
with realistic parametrization from
ref.~\cite{amb} at $Q^2=10~GeV^2$ (curve 1)
and $Q^2=90~GeV^2$ (curve 2). Dashed line is the same as
the solid line in fig.~1.
}\\[1cm]

%
%
%

Figure 3. {\em The the deuteron structure function $F_2^D$ as $x \to
1$.
Curves: BS approach (solid),
 non-relativistic $x$-rescaling
model (dashed) and  the free nucleon structure function (dotted).
The nucleon SF, $F_2^N$ is taken from
ref.~{\rm \cite{amb}} at $Q^2=50~GeV^2$.
}

\end{document}